\documentclass[12pt]{iopart}

\usepackage{graphicx}
\usepackage{epsfig}

\begin{document}

\title[Phase transition in the globalization of trade]{Phase transition in the globalization of trade}

\author{M. \'{A}ngeles Serrano}

\address{School of Informatics, Indiana University, Bloomington 47406,
IN, USA} \ead{mdserran@indiana.edu}
\begin{abstract}
Globalization processes interweave economic structures at a
worldwide scale, trade playing a central role as one of the
elemental channels of interaction among countries. Despite the
significance of such phenomena, measuring economic globalization
still remains an open problem. More quantitative treatments could
improve the understanding of globalization at the same time that
help a formal basis for comparative economic history. In this
letter, we investigate the time evolution of the statistical
properties of bilateral trade imbalances between countries in the
trade system. We measure their cumulative probability distribution
at different moments in time to discover a sudden transition circa
1960 from a regime where the distribution was always represented by
a steady characteristic function to a new state where the
distribution dilates as time goes on. This suggests that the rule
that was governing the statistical behavior of bilateral trade
imbalances until the 60's abruptly changed to a new form persistent
in the last decades. In the new regime, the figures for the
different years collapse into a universal master curve when rescaled
by the corresponding global gross domestic product value. This
coupling points to an increased interdependence of world economies
and its onset corresponds in time with the starting of the last
globalization wave.
\end{abstract}

\maketitle

\section{Globalization and trade}
At this point, the beginning of the third millennium in the
Gregorian calendar, it seems customarily assumed that several epochs
of globalization have occurred to shape the world as we perceive it
today. In these periods, a complex series of changes closely
intertwined develop into an increasing interdependence and
interaction between people and human organizations in disparate
locations of the world. These changes, when structural, are in a
great part of economic nature, markets becoming natural mediators of
globalization forces~\cite{Levitt:1983}.

In terms of trade~\cite{Krugman:1995}, there is still controversy
whether the last globalization waves, the first roughly identified
from 1870 till the beginning of World War I and the second from 1960
to the present, are more different than similar~\cite{Baldwin:1999}
--there is no a complete agreement either whether a third middle
wave has occurred~\cite{ChaseDunn:2000}. The two waves correspond to
processes of decolonization and falls of technical barriers with the
corresponding downloads of costs and time expenditures. The first
wave was triggered by the Industrial Revolution, with steam power
encouraging the expanding of railroad networks and oceanic routes
and the telegraph connecting the two sides of the Atlantic. The
second came intimately related to the Information Technologies
Revolution, communications costs dramatically dropping at the same
time that information management capabilities explode. In gross
terms, the first globalization burst had to do with lower costs in
transportation of materials and goods, while the second deals with
exchange of information and ideas. Two different natures of changes
that may well produce different impacts.

This discussion brings directly to the problem of how to measure
globalization. Quantitative approaches could enlighten but have been
timid to this moment, most works adopting analytical methodologies.
New quantitative ways of studying globalization processes are
required to exploit the wealth of information in historical
data~\cite{Maddison:2001,Maddison:2003}. One simple step forward
consists in complementing the study of aggregated or global values
with the statistical analysis of how they are distributed, from
where we can obtain not only more detailed but also new information.
For instance, a prominent figure when measuring globalization of
trade plots the evolution in time of total international trade as a
percentage of the global product, computed as the sum of all
national gross domestic products (GDPs). As it has been reported by
several authors~\cite{Baldwin:1999,ChaseDunn:2000,Taylor:2003}, an
{\it U-shape} pattern emerges, with rise of trade in the two eras of
globalization and a major reversal in between, and has been claimed
as a common trait which recurs in many other empirical analysis,
such as in the plot of global capital flows to GDP ratio or in the
correlation between savings and
investment~\cite{Baldwin:1999,Taylor:2005}.

\section{Analyzing the evolution of bilateral trade imbalances}
Still, further crucial information can be obtained from trade data
if, moreover, one draws the evolution in time of the distribution of
bilateral trade flows. Here we show that until 1960, all the
distributions overlap into a characteristic function, which
afterwards evolves widening as time goes by. Most interestingly, the
different trade imbalances distributions for all years since 1960
can be rescaled into a single master curve just by taking into
account the evolution in time of the global GDP, which marks a
characteristic scale with respect to which the system is
self-similar. The breaking of the original folding at the point when
a new single-curve collapse starts suggests that the rule that has
been governing the statistical behavior of bilateral trade
imbalances until the 60's has changed to a new scaling law in the
last decades, and that this has happened in a sudden transition of
the world trade system just at the beginning of the last
globalization wave.

\begin{figure}[t]
\begin{center}
  \epsfig{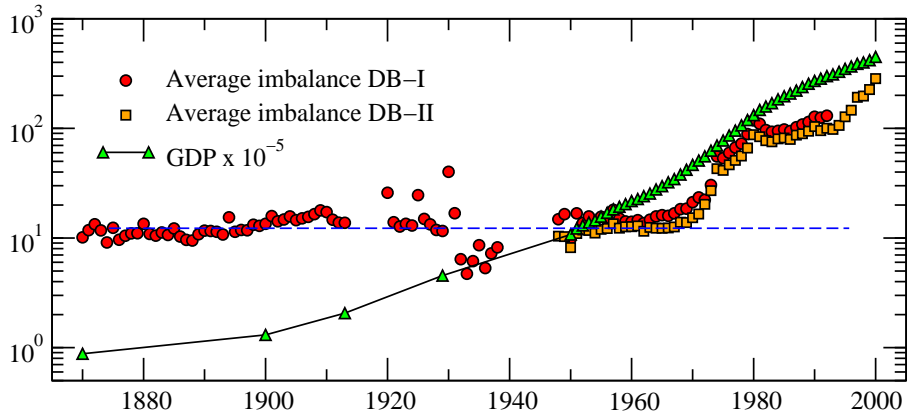}
\end{center}
  \caption{Historical evolution of global GDP values and average trade imbalances. From 1870 to 2000, the average value of the bilateral trade
  imbalances between countries in the world trade system for the two data bases that
  we are considering is compared to the
  evolution in the same period of the global gross domestic product (GDP).
  The results are shown on a log-linear scale for ease of comparison.
  The unit of measure for the average imbalance is millions of current year US dollars.
  GDP is also given in millions of current year US dollars, but it is rescaled to the level of bilateral trade
  imbalances.}
  \label{fig:1}
\end{figure}
We used historical national import/export data from two different
databases, DBI~\cite{Barbieri:2002,BarbieriData} (1870-1992) and
DBII~\cite{Gleditsch:2002,GleditschData} (1948-2000). World war
periods, 1914-1919 and 1939-1947, are avoided due to lack of
reported information in the data bases. World GDP historical values
are more difficult to obtain. Despite its role as a major instrument
of economic policy in virtually all countries in the world, it is
indeed a twentieth century concept as the rest of the national
income accounts and GDP data were not collected or even defined
before the 1930s. In our graphs, GDP data come from a third
source~\cite{Maddison:2007,MaddisonData}. The bilateral trade flux
$F$ between two countries is computed as the net money flow from one
to the other due to trade exchanges. In Fig.~\ref{fig:1}, the time
evolution of the global GDP is compared to that of the average
bilateral trade flow. Both evolutions seem decoupled until the 60's.
From 1870 to that date, the average imbalance remains fluctuating
around a constant level, in contrast to the estimation for the
global GDP, which appreciably grows. Afterwards, they couple and
follow similar growing patterns. This seems indicative of a change
of behavior or transition.

\begin{figure}[t]
\begin{center}
  \epsfig{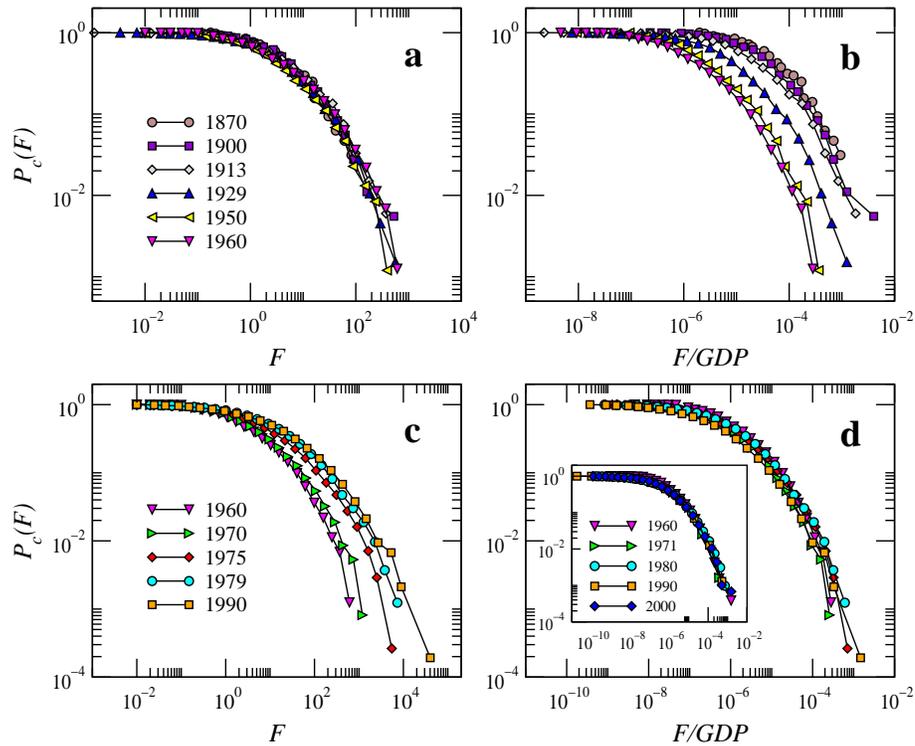}
  \caption{Master curve for the complementary cumulative probability
  distributions of bilateral trade imbalances. All the graphs
  are represented on a double logarithmic scale. The unit for all
  the variables is millions of current year US dollars. Complementary cumulative
  distributions of bilateral trade flows for several different years between 1870 and 1990
  are shown in {\bf(a)} and {\bf(c)}. Until 1960, all the
  distributions overlap on a characteristic curve {\bf(a)}. Afterwards,
  the distributions broaden as time goes on {\bf(c)}.
  In {\bf(b)} and {\bf(d)}, we show the rescaling of the previous graphs. The values on the $F$ axis has been divided by the GDP
  value for the corresponding year. The original characteristic curve
  prevailing before 1960 is broken {\bf(b)}, while a new master curve emerges from the collapse of the data between
  1960 and 1990 {\bf(d)}. The inset in {\bf(d)}
  shows that this collapse is also found when analyzing data from a different source which include the year 2000.}
  \label{fig:2}
\end{center}
\end{figure}
In Fig.~\ref{fig:2}, we present the complementary cumulative
probability distribution of net trade flows between pairs of
countries for several different years since 1870. The curves are
measuring the probability that a trade imbalance between two
countries in the trade system is bigger than a certain amount, and
the cumulative evaluation offers the advantage of filtering out the
statistical noise due to the finite sizes of the samples without
loosing information about the distribution. The first to observe is
that, in good approximation, all the curves overlap between the
years 1870 and 1960, see Fig.~\ref{fig:2}(a), moment in which the
folding is broken and the distributions evolve widening year after
year, see Fig.~\ref{fig:2}(c). We now rescale the curve for each
year by taking into account the global GDP value in that period. For
each distribution, we apply the transformation $F\longrightarrow
F/GDP$ which divides the fluxes in the horizontal F axis by the
corresponding global GDP value. Notice that this transformation
would produce the same result if real values were used instead of
nominal values. The results are shown in Fig.~\ref{fig:2}(b) and
(d). In Fig.~\ref{fig:2}(b), we just corroborate that the collapse
does not work for the years between 1870 to 1960, the transformation
indeed breaks the original folding. On the contrary, in
Fig.~\ref{fig:2}(d) all the distributions for different years since
1960 and until 2000 and for the two different databases under
consideration show an excellent collapse into a single master curve.
Imports and exports on their own are seen to present the same
behavior, due to the high correlation between the levels of imports
and exports in every single trade channel. All these distributions
are lognormal --they can be thought of as the multiplicative product
of many small independent factors--, again an ubiquitous shape in
economics. They can be adjusted to the form
$1/2-1/2*\mathrm{erf}[(\ln(x)-\mu)/(\sigma\sqrt{2})]$, where $\mu$
and $\sigma$ are the mean and standard deviation of $ln(x)$ and
$\mathrm{erf}$ stands for the error function. For the characteristic
curve in Fig.~\ref{fig:2}(a), $x\equiv F$ and the parameters are
$\mu=1.14$ and $\sigma=1.88$. The parameters for the master curve in
Fig.~\ref{fig:2}(d) with $x\equiv F/GDP$ are $\mu=-14.10$ and
$\sigma=2.34$, and for the inset $\mu=-14.39$ and $\sigma=2.55$, so
the two databases produce consistent information.

The collapse of the distribution for the different years into a
master curve implies that the system is self-similar with respect to
the characteristic scale given by the GDP. That means that the
widening of the distributions in time is just a dilatation driven by
the increase in total GDP and the same curve is found once the GDP
growth is discounted. Until new structural changes impact the world
trade system, we can assume that this behavior will be preserved
through time so that, by taking into account global GDP projection
values, we can predict the statistical distribution of bilateral
trade imbalances that would correspond to future periods.

\section{Discussion}
To understand better how GDP and trade imbalances are related, the
empirically successful gravity model of international trade can be
revisited~\cite{Bergstrand:1985}. In its basic form, it predicts
bilateral trade flows based on a functional form that is reminiscent
of the law of gravity in physics, and involves the distance between
pairs of countries and their economic masses, estimated in first
instance as their GDP. Thus, bilateral trade imbalances seem to be
empirically dictated in part by the GDP values of the countries
concerned. On the other hand, the aggregated value of all bilateral
trade flows for a certain country affects in its turn GDP levels.
The GDP of a country is defined as the market value of all final
goods and services produced within its borders in a given period of
time. In the expenditure-based approach, it is decomposed in several
terms as $GDP(t)=C(t)+I(t)+G(t)+F(t)$, where $C(t)$ stands for
private consumption, $I(t)$ for business investments in capital,
$G(t)$ for government spending, and $F(t)$ stands for net trade
balance. So, internal contributions are corrected by the trade
interactions with other countries. The sudden transition in the 60's
ties, from that moment, the evolution of the statistical
distribution of net trade flows to that of the global GDP. Although
not a proof, this seems to suggest that the internal components of
the economies (private, business and government spending) become
more dependent on trade exchanges with other countries, and this
leads us to conjecture that GDP and international trade, or in other
words internal and external components, are entangled in a complex
continuous feedback mechanism. At this stage, this idea is purely
speculative and further support should be provided in future work.
Nevertheless, the integration of markets in the last decades does
not seem to come in a smooth gradual transformation but rather in a
fault transition. Whether this is indicative of the birth of a truly
global market where all the economies are effectively interwoven
beyond trade needs more prove. New validations about the reach of
the phase transition should be performed as well. One example of
technical question that immediately arises is whether the empirical
success of gravity models for explaining bilateral trade as a
function of GDP after the 1960 is maintained when studying
historical data before this date.

In a way that is complementary to more traditional approaches in
economics, we measure a sudden transition in the statistical
behavior of trade imbalances which corresponds in time with the
starting of the last globalization wave. From a regime where the
distribution of trade imbalances was steady and independent of the
evolution of GDP, a new state is reached where the global GDP marks
the characteristic scale and the distribution dilates as GDP grows
in time. In the new regime, the distributions for the different
years collapse into a single universal master curve when rescaled by
the corresponding global GDP value. Although more work should be
done in order to clarify the relation between these empirical facts
and the effects of the last wave of globalization in trade, the
empirical findings in this paper point to an abrupt transformation
that has qualitatively and quantitatively increased the
interdependence of world economies through trade. From these results
we can conjecture that the segregation of factors in purely internal
components, just related to individual economies, versus trade,
implying interactions with other countries, seems not to be neat any
more. As a conclusion, the mere aggregation of bilateral trade
exchanges is not enough to explain the emergent behavior of the
global trade system, that, on the other hand, we can predict at the
statistical level. While we do not see any other radical revolutions
impacting the trade system, and maybe this is not so difficult, the
self-similar character of the distribution of bilateral trade
imbalances allows us to anticipate its form for the forthcoming
years.

\section*{Acknowledgments}
We thank Mari\'{a}n Bogu\~{n}\'{a} and Alessandro Vespignani for
helpful comments.

\section*{References}

\end{document}